\documentclass{emulateapj}

\usepackage{epsfig}
\usepackage{graphicx}
\usepackage{epstopdf}
\usepackage{morefloats}
\usepackage{amsmath}
\usepackage{natbib}
\usepackage{aas_macros}

\shorttitle{Analysis of CBe Star Axis Ratios}
\shortauthors{Cyr et al.}

\begin{document}

\title{Statistical Analysis of Interferometric Measurements of Axis Ratios for Classical Be Stars}
\author{ R. P. Cyr \altaffilmark{1}, C. E. Jones\altaffilmark{1},  C. Tycner\altaffilmark{2}}
\altaffiltext{1}{Department of Physics and Astronomy, Western University, London, ON Canada N6A 3K7}
\altaffiltext{2}{Department of Physics, Central Michigan University, Mt. Pleasant, MI 48859 USA}

\begin{abstract}

This work presents a novel method to estimate the effective opening angle of CBe star disks from projected axis ratio measurements, obtained by interferometry using Bayesian statistics.  A Monte Carlo scheme was used to generate a large set of theoretical axis ratios from disk models using different distributions of disk densities and opening angles. These theoretical samples were then compared to observational samples, using a two-sample Kolmogorov-Smirnov test, to determine which theoretical distribution best reproduces the observations. The results suggest that the observed ratio distributions in the K-, H-, and N-band can best be explained by the presence of thin disks, with opening half-angles of the order of 0.15$\degr$ to 4.0$\degr$. Results for measurements over the H$\alpha$ line point toward slightly thicker disks, 3.7$\degr$ to 14$\degr$, which is consistent with a flaring disk predicted by the viscous disk model.

\end{abstract}


\section{INTRODUCTION}
\label{intro}

Classical Be (CBe) stars are fast rotating, non-supergiant B-type stars surrounded by a thin gaseous Keplerian disk. The majority of the distinctive spectral features of CBe stars (such as Balmer emission lines, infrared excess, polarization) originates from this circumstellar envelope. It is now widely accepted that the disk does not form from infalling material, as is the case in accretion systems, but rather from outflowing material from the central star itself, in what is sometime referred to as a "decretion" system.

The structure of these circumstellar envelopes has been the subject of many studies. \citet{beh59} was the first to measure linear polarization in a CBe star, which gave the first evidence that the envelope might have a preferred orientation as opposed to a purely spherical shape. Later studies performed by \citet{dou92}, \citet{qui94a}, \citet{ste95}, and \citet{qui97}, to name a few, confirmed that the envelope was not spherical, but disk shaped. It is now widely accepted that the gas envelopes around CBe stars are in the shape of a thin disk. However, the question remains, exactly how thin are these disks?

The thickness of the equatorial disk is often defined in terms of the opening angle. Several studies attempted to estimate the opening angles of CBe star disks. By comparing the ratio of Be-shell stars to all CBe stars, \citet{por96} estimated the opening angles of 5$\degr$. However, using a similar method, \citet{han96} estimated an opening half-angle of 13$\degr$. From spectroscopic and interferometric measurements, \citet{qui97} estimated the upper limit of the opening half-angle for $\zeta$ Tau to be 20$\degr$, whereas \citet{woo97} estimated 2.5$\degr$ for the same star. 

In recent years, many groups have used optical interferometry to study the shape and extent of CBe star disks. One common measurement is the projected axis ratio of the disk, that is the ratio of the shortest to the longest axis as projected in the plane of the sky (i.e. the minor to major axis ratio). 

Measurements of axis ratios have been widely used, and still are, to investigate the geometry and extent of elliptical galaxies \citep{san70,lam92}, globular clusters \citep{fal83} as well as molecular cloud cores and bok globules \citep{ryd96, jon01, jon02}. Only recently has the number of available measurements of CBe star disk ratios been large enough to attempt such a study for these objects.

In this work, a new method of deprojecting the true shape distribution of CBe star disks using observed axis ratios is presented. We accomplish this by constructing a set of simulated observations, using disk models and various shape distributions, which we compare to actual observations using Bayesian statistics.

\section{OBSERVATIONAL DATA}
\label{sec:ob}

Interferometric instruments allow us to observe objects at a much smaller angular scale than conventional telescopes. Interferometry is therefore the perfect tool to study the shape and extent of CBe star disk. The current generation of interferometric instruments typically have an angular resolution of the order of milliarcsecond \citep{gie07, tyc11}. Despite this, the number of projected axis ratio measurements are still limited to CBe stars within a few hundred parsecs of the Earth.

It's important to note that interferometry does not measure axis ratios directly, as it is not possible to directly image the disks, but instead measures the visibility of the star/disk system for a given baseline. These measurements provide information on the projected extent of the disk along the axis parallel to the baseline used. Once enough measurements are acquired, models of the expected visibility curve for the CBe star system are then applied to the observations and their parameters are adjusted until a best fit is found. Typical free parameters of these models include angular size of the major and minor axis, from which the axis ratio can be calculated. More details on the methodology and models used can be found in the papers referred to at the bottom of Table~\ref{table:ratio}.

All axis ratio measurements used in this work were obtained through interferometric measurements and were gathered from the literature. Measurements were selected following certain criteria. Only measurements within certain wavelength regimes were considered, namely the K-band, H-band, N-band, and at the H$\alpha$ emission line. We also rejected axis ratios that were used as a fixed parameter within the model fitting, as those values were assumed prior to the measurements as opposed to deduced from them. Finally, we rejected measurements with very high level of uncertainties (typically those around or greater than 1.0).\footnote{Section~\ref{subsec:comp} discusses how the uncertainty is used to weight each measurement.}

Table~\ref{table:ratio} shows the resulting compilation of observed axis ratios. These ratios are ordered by the HR number of their corresponding star. Also provided in Table~\ref{table:ratio} are the common name for each star, the wavelength regime of the observation, and the reference for each observation. Note that some author(s) used two or more different models in order to fit the same measurements, resulting in two or more ratio values for the same star. In those cases, all ratio values were included, unless these models were explicitly rejected by the author(s) of that particular study.

\begin{deluxetable}{lcccc}
\tablewidth{0pt}
\tablecaption{Observed apparent axis ratios from literature.}
\tablehead{ \colhead{HR Number} & \colhead{Star Name} & \colhead{Wavelength} & \colhead{Ratio} & \colhead{Reference}\\
\  &  & \colhead{regime} &  & }
\startdata

HR 193 & o Cas & K-band & 0.58 $\pm$ 0.10 & 1 \\

HR 264 & $\gamma$ Cas & H$\alpha$ line & 0.70 $\pm$ 0.02 & 2 \\ 
& & & 0.77 $\pm$ 0.02 & 2 \\
& & & 0.79 $\pm$ 0.03 & 3 \\
& & & 0.58 $\pm$ 0.03 & 4 \\
& & H-band & 0.75 $\pm$ 0.05 & 5 \\ 
& & K-band & 0.59 $\pm$ 0.04 & 6 \\ 
& & & 0.72 $\pm$ 0.04 & 1 \\ 

HR 496 & $\phi$ Per & H$\alpha$ line & 0.46 $\pm$ 0.04 & 2 \\
& & & 0.47 $\pm$ 0.05 & 2 \\
& & & 0.27 $\pm$ 0.01 & 4 \\ 

HR 936 & $\beta$ Per & K-band & 0.75 $\pm$ 0.04 & 7 \\	

HR 1087 & $\psi$ Per & H$\alpha$ line & 0.35 $\pm$ 0.03 & 8 \\
& & & 0.47 $\pm$ 0.11 & 2 \\
& & & 0.54 $\pm$ 0.07 & 2 \\
& & & 0.33 $\pm$ 0.01 & 9 \\
& & K-band & 0.25 $\pm$ 0.56 & 1 \\

HR 1165 & $\eta$ Tau & H$\alpha$ line & 0.95 $\pm$ 0.22 & 2 \\ 
& & & 0.98 $\pm$ 0.06 & 2 \\
& & & 0.75 $\pm$ 0.05 & 3 \\
 
HR 1180 & 28 Tau & K-band & 0.74 $\pm$ 0.10 & 7 \\

HR 1273 & 48 Per & H$\alpha$ line & 0.76 $\pm$ 0.08 & 8 \\ 
& & & 0.86 $\pm$ 0.18 & 2  \\
& & & 0.89 $\pm$ 0.13 & 2 \\
& & & 0.71 $\pm$ 0.03 & 9 \\

HR 1910 & $\zeta$ Tau & H$\alpha$ line & 0.30 $\pm$ 0.03 & 10 \\
& & & 0.28 $\pm$ 0.02 & 2 \\ 
& & & 0.30 $\pm$ 0.02 & 2 \\ 
& & & 0.31 $\pm$ 0.07 & 11 \\
& & H-band & 0.24 $\pm$ 0.14 & 12 \\ 
& & K-band & 0.09 $\pm$ 0.22 & 6 \\
& & & 0.15 $\pm$ 0.03 & 1  \\

HR 2845 & $\beta$ CMi & H$\alpha$ line & 0.69 $\pm$ 0.15 & 3 \\
& & H-band & 0.76 $\pm$ 0.10 & 13 \\

HR 4830 & BZ Cru & K-band & 0.62 $\pm$ 0.01 & 14 \\
& & H-band & 0.64 $\pm$ 0.02 & 14 \\

HR 5938 & 4 Her & K-band & 0.27 $\pm$ 0.08 & 1 \\

HR 5941 & 48 Lib & H-band & 0.60 $\pm$ 0.11 & 15 \\

HR 5953 & $\delta$ Sco & H-band & 0.77 $\pm$ 0.21 & 16 \\ 

HR 6510 & $\alpha$ Ara & K-band & 0.37 $\pm$ 0.12 & 17 \\
& & N-band & 0.38 $\pm$ 0.18 & 17\\
& & & 0.42 $\pm$ 0.17 & 17\\

HR 6779 & o Her & K-band & 0.44 $\pm$ 0.28 & 1 \\

HR 7106 & $\beta$ Lyr & K-band & 0.60 $\pm$ 0.05 & 7 \\

HR 7763 & P Cyg & K-band & 0.85 $\pm$ 0.02 & 7 \\

HR 8146 & $\upsilon$ Cyg & K-band & 0.26 $\pm$ 0.13 & 1 \\
& & & 0.42 $\pm$ 0.30 & 7 \\

HR 8402 & o Aqr & K-band & 0.25 $\pm$ 0.06 & 1 \\

HR 8773 & $\beta$ Psc & K-band & 0.70 $\pm$ 0.15 & 1 \\ 

\enddata

\tablerefs{1. \citet{tou13}; 2. \citet{qui97}; 3. \citet{tyc05}; 4. \citet{tyc06}; 5. \citet{smi12}; 6. \citet{gie07}; 7. \citet{grz13}; 8. \citet{del11}; 9. Grzenia, B. J. (not yet published); 10. \citet{qui94b}; 11. \citet{tyc04}; 12. \citet{sch10}; 13. \citet{kra12}; 14. \citet{ste13}; 15. \citet{ste12}; 16. \citet{mil10}; 17. \citet{mei09}}
\label{table:ratio}
\end{deluxetable}

\section{THEORY}
\label{sec:theory}

\subsection{Viscous Disk Models}
\label{subsec:viscous}

At present, the viscous decretion disk model is the most widely accepted model to explain CBe star disk growth. It was first proposed by \citet{sha73} as a way to explain the inward flow of material in accretion disk systems, such as forming stars and black holes. The model was later modified \citep{lee91, por99, oka01} using the standard $\alpha$-prescription theory to include systems with outward flow of material, such as CBe stars. The model proposes that material from the equatorial region of the stellar atmosphere is injected at Keplerian orbital velocity into the base of the disk by some yet unknown mechanism.  If the material is steadily supplied by the star, it will start interacting with itself through a process referred to as viscosity, causing parts of the gas to slow down and settle into orbits close to the star, and other parts to spun up and move to greater radial distance from the star. Angular momentum will therefore be transferred from the star and carried outward into the disk. 

Further understanding of this model requires solving the hydrodynamic equations. We will not go through the derivations in this work, however, different approaches to solve these equations as well as their interpretations have been presented by various authors. \citet{car08} looked at the solution for a non-isothermal disk, while \citet{oka07} and \citet{hau11} described the solution for a system with a varying mass transfer rate. For this discussion, we will look at the results presented in \citet{car11}. Starting with some basic assumptions (no self gravity in the disk, slow radial velocity component, and a vertical structure in hydrostatic equilibrium), the following density structure equation is obtained:

\begin{equation}
\rho(r,z)=\rho_{0} r^{-n} \exp[-0.5(z/H)^2],
\label{eq:rho}
\end{equation}
where $\rho_0$ is the density at the base of the disk, $r$ and $z$ are the radial distance and height above the disk, respectively (both are expressed in stellar radii), $n$ is the power law describing how the density falls off, and $H$ is the scale height of the disk. The scale height $H(r)$ depends on the sound speed inside the disk ($c_s$) and the Keplerian velocity at the equator of the star ($V_{\text{Kep}}$): 
\begin{equation}
H(r) = \frac{c_s}{V_{\text{Kep}}}r^{3/2}.
\label{eq:h}
\end{equation}

\begin{figure}
\plotone{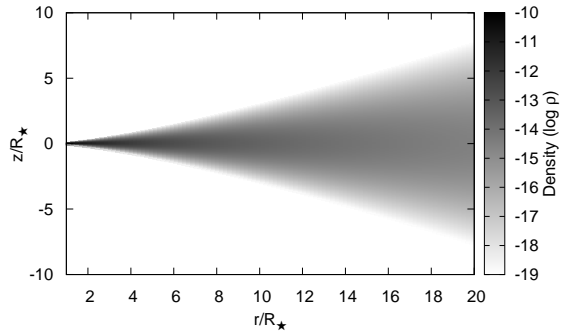}
\caption{Density structure of a typical CBe star disk with a base density of $\rho_0 = 10^{-10}$ g~cm$^{-3}$ and a fall off power law of $n = 3.5$, following Equation~(\ref{eq:rho}). The radial ($r$) and vertical ($z$) positions are expressed in units of stellar radii ($R_{\star}$) while the grayscale is in units of log($\rho$).}
\label{fig:density}
\end{figure}

Using these equations, \citet{car11} derives a value of 3.5 for $n$ in the case of an isothermal disk. However, values ranging from 2 to 5 have been found for IR observations \citep{wat87}, interferometric measurements \citep{tyc08, jon08}, and from the H$\alpha$ line profile modelling \citep{sil10}. As an example, Figure~\ref{fig:density} shows the density structure of a disk with $\rho_{0} = 10^{-11}$ g~cm$^{-3}$ and $n = 3.5$, based on Equations~(\ref{eq:rho}) and~(\ref{eq:h}).

\subsection{Geometry of Disk Models}
\label{subsec:model}

Two basic disk shapes are used as models to describe the disks of CBe stars in this investigation. For simplicity, both models assume no irregularities in the disk, meaning that we have an azimuthal symmetry (axisymmetric disks) and a symmetry above and below the plane of the disk (longitudinal symmetry).  

\begin{figure}
\plotone{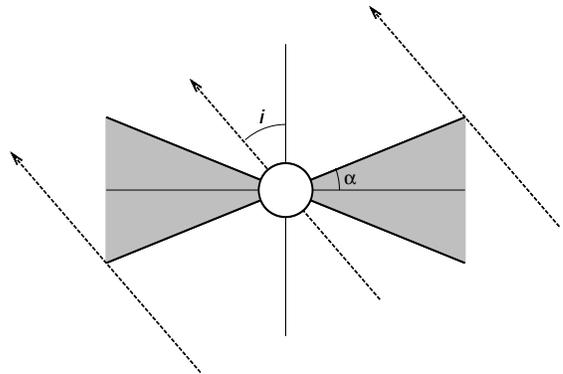}
\caption{Cross-sectional view of the wedge shape disk model. The grey area represents the material of the disk. The dashed arrows point toward the line of sight of the observer.}
\label{fig:wedge}
\end{figure}

The first model consists very simply of a disk whose scale height increases linearly with radius, leading to a wedge geometry when seen as a cross-section, similar to the one proposed by \citet{wat86}. The disk is truncated at a certain radius $R$ greater than the stellar radius $R_{\star}$, and the rate of vertical increase is defined by the opening half-angle parameter ($\alpha$) which is tied to the shape of the disk. For simplicity, it is also assumed the disk is completely opaque inside the wedge. Figure~\ref{fig:wedge} shows a cross-section of this model.

The projected axis ratio ($q$) for this model is a function of two parameters; $\alpha$ and the inclination angle $i$. Considering this wedge shape disk model truncated at $R$, with an opening half-angle $\alpha$ and is observed at $i$. To the observer, the projected length of the axis perpendicular to the plane of the inclination (perpendicular to the page in Figure~\ref{fig:wedge}) will not be affected by the inclination, and will simply be the length of the equatorial disk,\footnote{This is of course not the angular dimension, as we did not scale for distance. However, for the purpose of this work, we are only interested in the ratio of the axes.} 
\begin {equation}
 L_{1}=2R.
\label{l1}
\end {equation}

The projected length of the axis in the plane of the inclination (the cross-section depicted in Figure~\ref{fig:wedge}), however, will be the most affected by the inclination angle. As Figure~\ref{fig:wedge} shows, the projected length of this axis is along the plane of the sky between the lower edge of the disk facing toward the observer and the upper edge of the disk, facing away from the observer. Its measured length, for inclinations between $i = 0\degr$ and $i = 90\degr$, is given by
\begin {equation}
 L_{2} =2R \frac{\cos(i-\alpha)}{\cos(\alpha)}.
\label{l2}
\end {equation}
The ratio of $L_{1}$ and $L_{2}$ two is therefore our theoretical projected axis ratio and is given by
\begin {equation}
 q' (i,\alpha)=\frac{\cos(i-\alpha)}{\cos(\alpha)}.
\label{rprime}
\end {equation}

Note that the dependence on the $R$ is gone, leaving only $\alpha$ and $i$ as variables. To ensure that the ratio is always between 0 and 1 (minor axis over major axis) the equation 
\begin {equation}
 q (i,\alpha)=1-\vert 1-q'(i,\alpha) \vert
\label{eq_r}
\end {equation}
is used.
 
One advantage of this model is that our axis ratio calculations depend only on these two parameters. However, the shape of this model is very simplistic, and does not take into account the density and thermal structure nor the optical thickness of the disk, all of which play a role in the emission processes.
 
\begin{figure}
\plotone{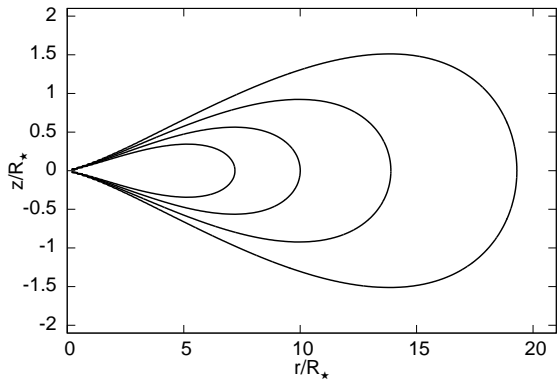}
\caption{Cross-sectional view of the equidensity (ED) disk model for different density ratios ($\wp$). From the smallest shape to the biggest, the $\wp$ values are $10^{-3.0}$, $10^{-3.5}$, $10^{-4.0}$, and $10^{-4.5}$.}
\label{fig:equidensity}
\end{figure}

For the second model, we chose a shape that is more related to the density structure of CBe star disk, as predicted by the viscous disk model. As mentioned above, the emission is related to disk density. We therefore decided to base our model on the shape of the equidensity (ED) regions of the disk; that is, the shape of the regions where the density is uniform. Once again, we did not take into account the thermal structure, assuming therefore that we have an isothermal disk. By keeping $\rho$ constant, Equation~(\ref{eq:rho}) becomes an implicit function of $r$ and $z$:
\begin{equation}
\wp =r^{-n} \exp \left [ -\frac{1}{2}\left ( \frac{v z}{r^{1.5}} \right )^2 \right ],
\label{eq:z}
\end{equation}
where $\wp = \rho / \rho_0$ is the ratio of the density of the region versus the density at base of the disk, and $v$ is the $V_{\text{Kep}}$ over the $c_s$ ratio. These two parameters, along with \textit{n} are the physical parameters of this model. Isolating $z$ from Equation~(\ref{eq:z}) gives the relationship between $z$ and $r$:
\begin{equation}
z^2=-\frac{2 r^3}{v^2} \ln(\wp r^{n}),
\label{eq:z2}
\end{equation}
or,
\begin{equation}
z(r)=\pm \frac{\sqrt{-2 r^3 \ln(\wp r^{n})}}{v}.
\label{eq:z(r)}
\end{equation}
Figure~\ref{fig:equidensity} shows various ED regions for different values of $\wp$, using the same disk parameters as Figure~\ref{fig:density}. 

Equation~(\ref{eq:z(r)}), however, is a rather complex function, the complexity of which increases dramatically when taking its derivative with respect to $r$ (which, as discussed later, is required to calculate the projected axis ratio). We therefore decided to use a simpler function that can reproduce ED shape given by Equation~(\ref{eq:z(r)}). This function is described as follows:
\begin{equation}
z^2 = a r^2 (1 - b r^2),
\label{eq:lemniscate}
\end{equation}
and the resulting shape is called a lemniscate, where $a$ and $b$ are its shape parameters. Equation~(\ref{eq:lemniscate}) can be rewritten as a function by isolating $z$:
\begin{equation}
z(r) = \pm r \sqrt{a - a b r^2}.
\label{eq:z_func}
\end{equation}

In order to best reproduce the ED profile with Equation~(\ref{eq:lemniscate}), the shape parameters $a$ and $b$ have been associated with the three physical parameters of Equation~(\ref{eq:z}), namely $\wp$, $n$, and $v$. Let's define $r_{\text{max}}$ and $z_{\text{max}}$ as the maximum radial and vertical extent, respectively, of our shape, and $r_z$ such that $z(r_z) = z_{\text{max}}$, i.e. the radial position where the vertical extent is maximum. The ED and lemniscate shapes are differentiated by the use of the superscripts $E$ and $L$, respectively.  As seen in Figure~\ref{fig:equidensity}, $z=0$ at $r_{\text{max}}$. Setting $z$ to 0 in Equations~(\ref{eq:z(r)}) and (\ref{eq:z_func}), and solving for $r$ gives a maximum radial position for both shapes \footnote{Multiple solutions exists for Equation~(\ref{eq:z_func}) but are rejected as only the maximum positive value is sought.}:
\begin{equation}
r_{\text{max}}^E=\wp^{-1/n},
\label{eq:rmax_ED}
\end{equation}
\begin{equation}
r_{\text{max}}^L=b^{-1/2}.
\label{eq:rmax_lem}
\end{equation}

Equating $r_{\text{max}}^E$ to $r_{\text{max}}^L$ gives the following expression for $b$:  
\begin{equation}
b=\wp^{2/n}.
\label{eq:param_b}
\end{equation}

Equation~\ref{eq:param_b} shows that the parameter \textit{b} and by extension $\wp$ and \textit{n} are the only parameters responsible for the horizontal extent of the disk.

The maximum $z$ extent will occur where the first derivative of $z$, with respect to $r$, is 0. Therefore, $r_z$ can be determined by setting $\mathrm{d}z/\mathrm{d}r = 0$ and solving for $r$. We note here that, as the maxima of $z^2$ occurs at the same radial position as $z$, $\mathrm{d}z^2/\mathrm{d}r = 0$ was used instead to simplify the expressions. Applying this procedure to Equations~(\ref{eq:z2}) (for the ED shape) and (\ref{eq:lemniscate}) (for the lemniscate shape) yields
\begin{equation}
\begin{split}
r_{z}^E &= \mathrm{e}^{-1/3} \wp^{-1/n}\\
&\approx 0.717 r_{\text{max}}^E,
\label{eq:rz_ED}
\end{split}
\end{equation}
\begin{equation}
\begin{split}
r_{z}^L &= 2^{-1/2} b^{-1/2}\\
&\approx 0.707 r_{\text{max}}^L,
\label{eq:rz_lem}
\end{split}
\end{equation}
respectively. We notice that, in both cases, the ratio of $r_z$ over $r_{\text{max}}$ is a constant value, without any dependence on the shape parameters. The fact that these constants differ for each shape indicates that it is impossible to match both $r_{\text{max}}$ and $r_z$ at the same time. Luckily, these constants differ very little from one another (less than 1.5$\%$), therefore the peaks can be said to be approximately at the same position.

An expression for $z_{\text{max}}^2$ can now be developed for both shapes by inserting Equations~(\ref{eq:rz_ED}) and (\ref{eq:rz_lem}) into Equations~(\ref{eq:z2}) and (\ref{eq:lemniscate}), respectively,
\begin{equation}
\left( z_{\text{max}}^E \right)^2 = \frac{2}{3 e} \frac{n}{v^2 P^{3/n}},
\label{eq:zmax_ED}
\end{equation}
\begin{equation}
\left( z_{\text{max}}^L \right)^2 = \frac{a}{4 b}.
\label{eq:zmax_lem}
\end{equation}
Finally, the shape parameter $a$ can be determined by setting $(z_{\text{max}}^E)^2 = (z_{\text{max}}^L)^2$ and substituting the expression for $b$ from Equation~(\ref{eq:param_b}):
\begin{equation}
\begin{split}
a &= \frac{8}{3 e} \frac{n}{v^2 P^{1/n}}\\
  &\approx \frac{n}{v^2 P^{1/n}}.
\label{eq:param_a}
\end{split}
\end{equation}

\begin{figure}
\plotone{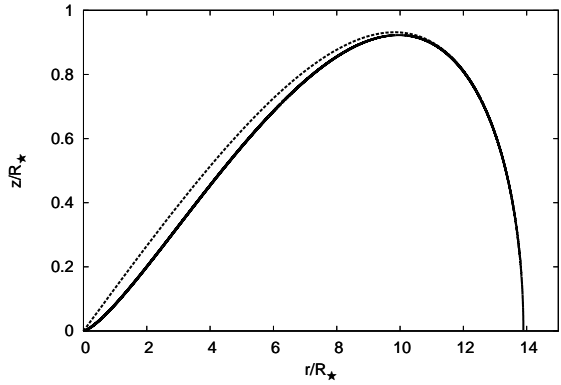}
\caption{Comparison of the equidensity shapes from the density equation (solid line) and the lemniscate equation (dashed line).}
\label{fig:lem}
\end{figure}

Figure~\ref{fig:lem} shows the ED curve as defined by the viscous disk model (solid line) and the lemniscate curve (dashed line), both using the same physical parameters ($n$, $\wp$, and $v$). We see that the lemniscate curve reproduces the ED curve well at the outer edge, that is for $r \geq r_z$, but not so well for $r < r_z$. This, however, is unimportant for the purposes of this work as only the outer region is needed to calculate the projected axis ratio, as will be demonstrated below.   

\begin{figure}
\plotone{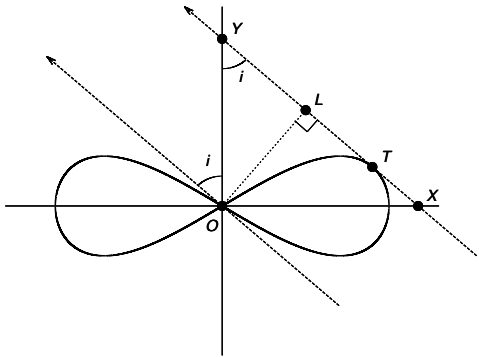}
\caption{Cross-sectional view of the equidensity disk model. The dashed arrows point toward the line of sight of the observer while the dotted line represents the projected size of the minor axis.}
\label{fig:lem_diag}
\end{figure}

To calculate the projected axis ratio, we must first determine the length of the major and minor axes for any $i$. To simplify things, we will only consider the ``half-length" instead of the full length, as the disk model is assumed to be axisymmetric. We will also only consider \textit{i} values between 0$\degr$ and 90$\degr$, again because of the symmetry of the disk. Figure~\ref{fig:lem_diag} shows a cross-section of our model disk. The dashed lines represent the lines of sight of an observer viewing the disk at an inclination angle $i$. 

When projected onto the plane of the sky, from the point of view of the observer, the extent of the major axis is equal to the dimension of the disk perpendicular to the inclination plane (perpendicular to the page in Figure~\ref{fig:lem_diag}), which in our case is the radius of the disk. This radius can easily be obtained from Equation~(\ref{eq:lemniscate}) by setting $z = 0$:
\begin{equation}
\begin{split}
L_{\text{major}} = r_{\text{max}} &= b^{-1/2}\\ 
 &= \wp^{-1/n}.
\label{eq:r_max}
\end{split}
\end{equation}

The extent of the minor axis is the projected dimension of the component of the disk in the inclination plane (plane of the page in Figure~\ref{fig:lem_diag}). Therefore, the dimension of the minor axis is the projection of the line segment $\overline{OT}$ (point $T$ being the point at which the line of sight intersects the disk tangentially) on the plane of the sky, which corresponds to  $\left | \overline{OL} \right |$ in Figure~\ref{fig:lem_diag}. It is easy to show that $\left | \overline{OL} \right |$ can be obtained from $\left | \overline{OX} \right |$ or $\left | \overline{OY} \right |$ by simple trigonometry. To obtain these points, we must first determine the position of point $T$. 

As mentioned earlier, $T$ is the point on the curve whose slope corresponds to the line of sight, in other words where $\frac{dz(r)}{dr} = \cot(i)$. The radial and vertical coordinates at point \textit{T} will be referred to as $r_{T}$ and $z_{T}$, respectively. Setting the derivative of Equation~(\ref{eq:z_func}) to $\cot(i)$ and solving for $r$ gives us the $r_{T}$ coordinate:
\begin{equation}
r_T = \sqrt{\frac{4a - \cot^2(i) + \sqrt{8a\cot^2(i) + \cot^4(i)}}{8ab}}.
\label{eq:r_t}
\end{equation}
We can then obtain $z_T$ by substituting Equation~(\ref{eq:r_t}) back into Equation~(\ref{eq:z_func}):
\begin{equation}
z_T =  r_T \sqrt{a - a b r_T^2}.
\label{eq:z_t}
\end{equation}
Either $\left | \overline{OX} \right |$ or $\left | \overline{OY} \right |$ can now be obtained from the linear equation using $r_T$ and $z_T$ as coordinates and $-\cot(i)$ as the slope:
\begin{equation}
\begin{split}
\left | \overline{OX} \right | = r_T + z_T \tan(i),\\
\left | \overline{OY} \right | =  r_T \cot(i) + z_T.
\end{split}
\label{eq:AB}
\end{equation}
The minor axis can now be calculated using simple trigonometry and either of the above equations:
\begin{equation}
\begin{split}
L_{\text{minor}} &= \left | \overline{OL} \right |\\ 
 &= r_T\,\cos(i) + z_T\,\sin(i).
\label{eq:L_min}
\end{split}
\end{equation}
Finally, the projected axis ratio can be calculated by taking the ratio $L_{\text{minor}}/L_{\text{major}}$:
\begin{equation}
\begin{split}
\frac{L_{\text{minor}}}{L_{\text{major}}} &= \frac{r_T\,\cos(i) + z_T\,\sin(i)}{\wp^{-1/n}}\\
&= \frac{r_T\,\cos(i) + z_T\,\sin(i)}{b^{-1/2}}.
\label{eq:ratio2}
\end{split}
\end{equation}

Although this last equation appears to depend on both parameters $a$ and $b$, it in fact only depends on the former. This can easily be shown by considering that both $r_T$ and $z_T$ are proportional to $b^{-1/2}$. In other words the projected axis ratio does not depend on the actual size of the disk, leaving $a$ as the fundamental shape parameter. For this model we chose to use $\chi$, defined as the ratio $z_{max}/r_{max}$, as the shape parameter of this model. By combining Equations~(\ref{eq:rz_lem}) and (\ref{eq:zmax_lem}) it can easily be shown that the ratio $z_{max}/r_{max}$ depends only on $a$ and that it can therefore be used as a fundamental shape parameter. For simplicity, we shall call this parameter $\chi$:

\begin{equation}
\begin{split}
\chi &\equiv \frac{z_{max}}{r_{max}} = \frac{1}{2} \sqrt{a}\\
&= \frac{1}{2v}\sqrt{\frac{n}{\wp^{1/n}}}.
\label{eq:zm_rm}
\end{split}
\end{equation}

The advantage of this model is that the shape used is closely related to the viscous disk model, which as already discussed is widely accepted as the mechanism responsible for disk growth and therefore dictates the shape of the density distribution of CBe disks. 

\subsection{Distribution of Shape Parameters}
\label{subsec:dist}

As discussed above, each model has a set of parameters that define the shape of the disk. Our goal is to find which shape parameter value, or range of values, best reproduces the observed axis ratios.

The inclination parameter $i$ is present in both models. Assuming no preferred inclination in our population of stars, we distribute our angles using $i = \cos^{-1}(u)$, where $u$ is selected from a uniform distribution such as $u \in [0,1]$, as it was done by \citet{cra05}. As we assume the disk is symmetric in both models, the range of $i$ is limited to [0,$\pi/2$].

A $\beta$-distribution function was used to study the distribution of shape parameters. This distribution was chosen because it is well defined and well constrained within a finite interval. Its functional form is:
\begin{equation}
\Psi_{\beta} (x; A, B) = \frac{x^{A-1}(1-x)^{B-1}}{\beta(A, B)},
\label{eq:betaDist}
\end{equation}
where $x$ is a continuous variable between 0 and 1, $A$ and $B$ are the shape parameters and $\beta(A, B)$ is the Beta function: 
\begin{equation}
\beta(A, B) = \int_{0}^{1}t^{A-1}(1-t)^{B-1} \mathrm{d}t.
\label{eq:betaFunc}
\end{equation}

The shape parameters, $A$ and $B$, affect the width of the distribution and their relative values affect the position of the peak. The greater $B$ is compared to $A$, the closer the peak will be to the lower $x$ values, and vice-versa. The distribution will be centred if $A = B$. Higher values of either parameter results in a smaller deviation (thinner distribution). The $\beta$-distribution can also be parametrized in terms of the parameters $\mu$ and $\nu$ by using the following relations:
\begin{equation}
\begin{split}
A &= \mu \nu,\\
B &= (1 - \mu) \nu,
\label{eq:mu_nu_A}
\end{split}
\end{equation}
where $0 \le \mu \le 1$ and $\nu > 0$. The parameter $\mu$ in this case represents the mean value of the distribution. The variance of the distribution can also be expressed as a function of these parameters:
\begin{equation}
\Psi_{var} = \frac{\mu (1 - \mu)}{1 + \nu}.
\label{eq:variance_beta}
\end{equation}

\subsection{Ratio Simulations and Comparison with Observations}
\label{subsec:comp}

Sets of simulated axis ratios are generated using Monte Carlo techniques. Each set consists of $10^6$ simulated projected axis ratios using a specific model (Section~\ref{subsec:model}) whose shape parameters are chosen randomly for a specific distribution (Section~\ref{subsec:dist}). The parameters of the chosen distributions are varied systematically from set to set, allowing us to assess which distribution best reproduces the observations.    

The observed ratios were grouped into sets according to the wavelength regime at which they were measured. The first set includes all measurements in the K-band, which includes 18 ratio measurements of 16 distinct stars. In an attempt to increase the amount of data points, all measurements taken in either the K-, H-, or N-band (from now on referred to as KHN-band) were also grouped together. The reason for this grouping is that emission in these three bands are likely formed within similar volumes of the disk \citep{car11} and therefore $\alpha$ should be similar for all three bands. Adding the observations in the H- and N-band adds five measurements and one new star, for a total of 24 ratios for 17 stars. The third and final set consists of observations acquired over the H$\alpha$ emission line (656.3 nm). This set contains 20 measurements, which includes seven distinct stars.

As seen in Table~\ref{table:ratio} of Section~\ref{sec:ob} some stars have multiple ratio measurements and in some cases in the same wavelength regime. For this reason, a weighted average is used to reduce each of these multiple measurements to a single value. First, an inverse-variance weighting is applied to the measurements:
\begin {equation}
 w^\prime_j = \frac{1}{\sigma_j^2},
\label{eq:weight}
\end {equation}
where $\sigma_j$ is the uncertainty of the measurement. The weights for each individual star are then normalized in such a way that their sums equal 1.

The degree of agreement between the observed and simulated projected axis ratios is determined using the two-sample Kolmogorov-Smirnov (K-S) test. This test compares the cumulative distribution function of our samples, both observed and simulated, and determines whether the null hypothesis (i.e. that both samples come from the same distribution) can be rejected or not. It can also be used as a ``goodness-of-fit" test to determine which distribution of simulated ratios best fit the observed distribution. The K-S statistic $D$ is defined as the largest difference between the cumulative distribution functions (CDFs) of the two samples being compared, $F_{1}(x)$ and $F_{2}(x)$ \citep[see][for further details]{num_rec07}:
\begin{equation}
D=\sup_{x}|F_{1}(x) - F_{2}(x)|.
\label{eq:KS1}
\end{equation}
The significance level of $D$ can be estimated by the following function:
\begin{equation}
P_{D}\approx Q_{\text{KS}}(\lambda) = 2 \sum_{j=1}^{\infty}(-1)^{j-1}\mathrm{e}^{-2j^2\lambda^2},
\label{eq:KS2}
\end{equation}
where 
\begin{equation}
 \lambda = D \left( \sqrt{N} + 0.12 + 0.11/\sqrt{N} \right),
\label{eq:KS3}
\end{equation}
and $N$ is the effective number of data points, derived from the number of data points in each sample ($n_1$ and $n_2$), 
\begin{equation}
N = \frac{n_1 n_2}{n_1+n_2}.
\label{eq:N}
\end{equation}

The null hypothesis can be rejected (i.e. the samples do not come from the same distribution) if $P_D$ is below the significance level $\alpha$. For this work, we used a significance level of 10\% ($\alpha=0.1$).

\section{RESULTS}
\label{results}

\subsection{Wedge Model}
\label{sub:wedge_res}

\begin{figure}
\plotone{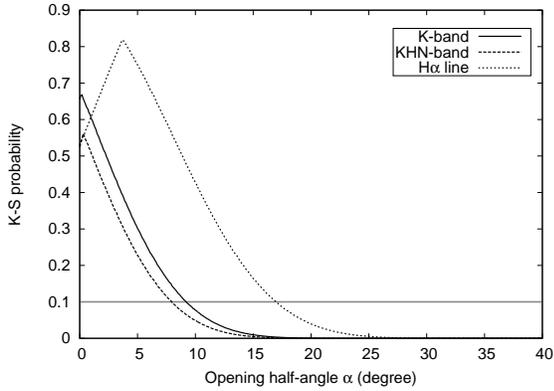}
\caption{K-S test results for simulated data with single $\alpha$ values for measurements in the K-band (solid line), the KHN-band (dashed line), and the H$\alpha$ line (dotted line). The grey line represents the 90$\%$ confidence limit of the test.}
\label{fig:singleAngle}
\end{figure}

We first compared the observed ratios with the simulated ratios obtained using single values of $\alpha$, that is no distribution was used. Figure~\ref{fig:singleAngle} shows the results of the K-S test between the observed and simulated ratios as a function of $\alpha$, for all three observational sets; K-band (solid), KHN-band (dash), and H$\alpha$ line (dot). The grey line represents the 90$\%$ confidence limit for this test.

For the K- and KHN-band, the model best reproduces the observations when small $\alpha$ values are used, that is for $\alpha$ of 0.15$\degr$ and 0.32$\degr$, respectively. After these maxima, the probability goes down exponentially, reaching the confidence limit at 9.21$\degr$ for the K-band and 7.94$\degr$ for the KHN-band. For H$\alpha$ observations, the model best matches the observations at $\alpha = 3.7\degr$, a value higher than the $\alpha$ found for the other two sets of observations. We also have a greater range of $\alpha$ within the confidence limit, which is reached at 28$\degr$.
 
Although these single-$\alpha$ simulations give us a good idea of the opening angle of our disks based on this simple assumed geometry, it is more probable that the opening angle of these disks are not all the same, but rather distributed over a certain range of angles. In order to take this into account, we repeated our simulations with $\alpha$ values randomly picked following a $\beta$-distribution (see Section~\ref{subsec:dist}) with different pairs of distribution parameters, $\mu$ and $\nu$, for each set of simulated ratios. The domain of the $\beta$-distribution, which is typically [0,1], was extended to match the range of $\alpha$. Once again, these simulated sets were compared with the observed ratios.   

\begin{figure}
\plotone{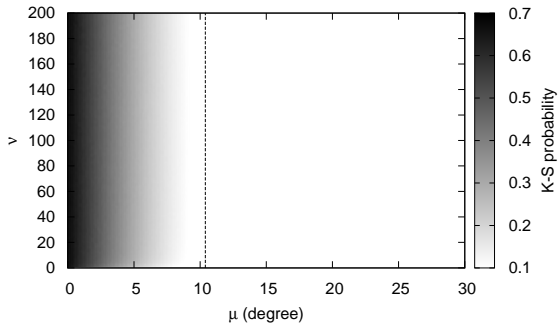}
\caption{Results of the K-S comparison test between the K-band set and the $\beta$-distributed simulation, using the wedge model, as a function of the $\mu$ and $\nu$ parameters. The dashed line shows the contour of the 90$\%$ confidence limit.} 
\label{fig:map_K}
\end{figure}

\begin{figure}
\plotone{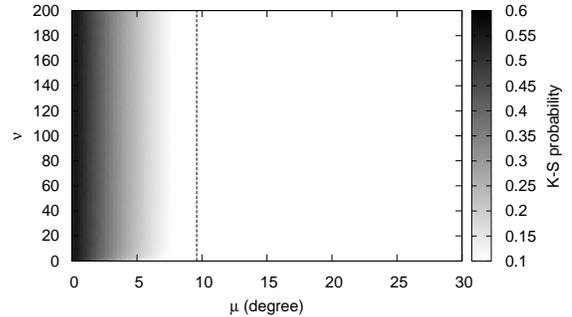}
\caption{Same as Figure~\ref{fig:map_K} except for the KHN-band set.} 
\label{fig:map_KHN}
\end{figure}

\begin{figure}
\plotone{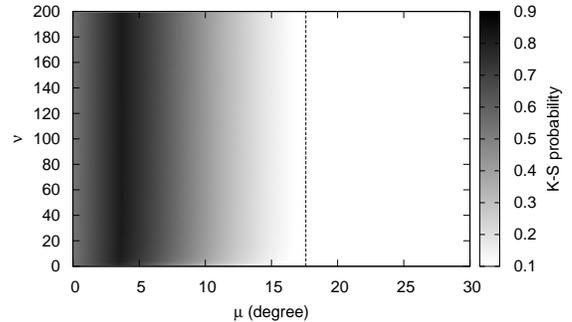}
\caption{Same as Figure~\ref{fig:map_K} except for the H$\alpha$ set.} 
\label{fig:map_Ha}
\end{figure}

Figures~\ref{fig:map_K}, \ref{fig:map_KHN}, and \ref{fig:map_Ha} show results of the K-S test for the K-band, KHN-band, and H$\alpha$ sets, respectively, as a function of the beta distribution parameters $\mu$ and $\nu$. The dash line shows the contour of the 90$\%$ confidence limit.

All three figures show a similar trend. The results of the K-S test seems to vary with $\mu$ but not, or at least not significantly, with $\nu$. This would indicate that the goodness of the fit depends almost entirely on the mean value of distribution of $\alpha$ but not its variance. For both the K- and KHN-band, the highest values of the K-S test results are located at low $\mu$ values, corresponding to $\beta$-distributions greatly skewed toward low $\alpha$ values and therefore thinner disks. For the H$\alpha$ set, the best fits appear at somewhat higher $\mu$ values, which correspond to a larger $\alpha$ (thicker disk). Interestingly, the highest K-S test results for all three sets occur at same mean $\alpha$ ($\mu$) values as the results of the previous test; 0.15$\degr$ for the K-band set, 0.32$\degr$ for the KHN-band set, and 3.7$\degr$ of the H$\alpha$ set.

The difference in the distributions of $\alpha$ from the K- and KHN-band sets, and the H$\alpha$ set is expected. \citet{car11}, have estimated that emission in the K-band, as well as in the H- and N-band, are formed in a much smaller volume of the disk near the star than the H$\alpha$ emission. Moreover, as discussed in Section~\ref{subsec:viscous}, the viscous disk model predicts a flaring of the disk, meaning that we expect the effective opening angle to be greater farther away from the star.

\subsection{Equidensity Model}
\label{sub:equidensity}

\begin{figure}
\plotone{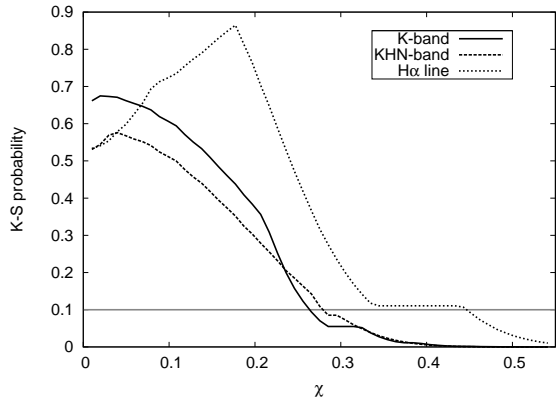}
\caption{K-S test results for simulated data with single $\chi$ values distribution for measurements in the K-band (solid line), the KHN-band (dashed line), and the H$\alpha$ line (dotted line). The grey line represents the 90$\%$ confidence limit of the test.} 
\label{fig:single_chi}
\end{figure}

Like the previous model, we start by comparing the observed ratios with simulated ratios obtained using single values for our shape parameters, which for this model is $\chi$. Figure~\ref{fig:single_chi} shows the results of the K-S test between the observed and simulated ratios as a function of $\chi$. Once again, the K-band, KHN-band, and H$\alpha$ line sets are represented by the solid, dashed, and dotted lines, respectively, while the grey line represents the 90$\%$ confidence limit.

For the K- and KHN-band, the best fit occurs at small values of $\chi$, that is 0.024 and 0.037, respectively, while the best fit for the H$\alpha$ set occurs for $\chi$ = 0.18, a value higher than the previous two. The range of $\chi$ including the confidence limit is also smaller for the K- and KHN-band (from 0 to 0.26 and 0 to 0.28, respectively) than H$\alpha$ (from 0 to 0.45). One notable feature is the small plateaus found near the end of each curve. These features are a result of the absence, in our samples, of observed axis ratio ranging somewhere between 0.25 and 0.45 depending on the observation set, which causes the maximum deviation between the observed and simulated CDFs to take similar values for a certain range of $\chi$ values.

\begin{figure}
\plotone{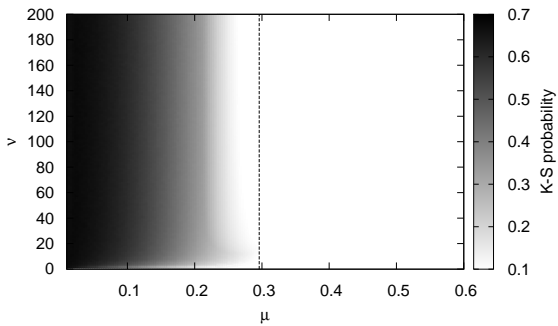}
\caption{Results of the K-S comparison test between the K-band set and the $\beta$-distributed simulation, using the ED model, as a function of the $\mu$ and $\nu$ parameters. The dashed line shows the contour of the 90$\%$ confidence limit.} 
\label{fig:map_K_rho}
\end{figure}

\begin{figure}
\plotone{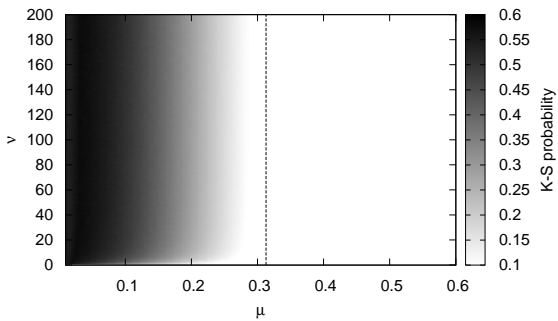}
\caption{Same as Figure~\ref{fig:map_K_rho} but for KHN-band set.} 
\label{fig:map_KHN_rho}
\end{figure}

\begin{figure}
\plotone{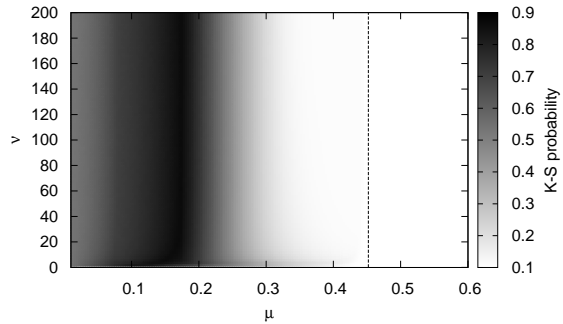}
\caption{Same as Figure~\ref{fig:map_K_rho} but for H$\alpha$ set.} 
\label{fig:map_Ha_rho}
\end{figure}

For the next step $\chi$ was varied over a $\beta$-distribution. The results of the K-S test using the equidensity model are presented in Figures~\ref{fig:map_K_rho}, \ref{fig:map_KHN_rho}, and \ref{fig:map_Ha_rho} for the K-band, KHN-band, and H$\alpha$ sets, respectively, in the same fashion as Figures~\ref{fig:map_K}, \ref{fig:map_KHN}, and \ref{fig:map_Ha} in the previous section.

Similarly to the results presented in the previous section for the wedge model, the K-S test results vary primarily with $\mu$ and are mostly independent of $\nu$. An exception to this can however be seen in the lower part of Figure~\ref{fig:map_Ha} for $\nu < 10$. Once again we see that the model better fits the K- and KHN-band observations when lower $\chi$ are used, while the H$\alpha$ line observations are best fit with a higher $\chi$. These results, as well as the results of the single $\chi$ simulation, agree with the results of the previous model. Assuming constant \textit{n} and \textit{v} values, Equation~(\ref{eq:zm_rm}) shows that lower $\chi$'s corresponds to a higher density ratio ($\wp$). As we can see from the density structure of the viscous disk (Figure~\ref{fig:equidensity}), ED regions with lower densities extend further away from the star and appear more "puffed up" vertically than regions of higher density. This means that these low density ED regions have a higher effective opening angle. Therefore, according to the equidensity model results, H$\alpha$ line comes from regions with a greater vertical extent, and therefore greater effective opening angles, than the KHN-band emitting regions.

To better compare the results of both models, we expressed the results of the ED model in terms of the effective opening half-angle, $\alpha_{\text{eff}}$. We define $\alpha_{\text{eff}}$ to be the angle between the base of the disk and the line going from the origin to the highest vertical point. The coordinates of this point are $r_{\text{z}}$ and $z_{\text{max}}$, as defined in Equations~(\ref{eq:rz_lem}) and (\ref{eq:zmax_lem}), respectively. The values of $\alpha_{\text{eff}}$ can then be obtained using simple trigonometry;
\begin{equation}
\alpha_{\text{eff}} = \arctan\left(\frac{z_{\text{max}}}{r_{\text{z}}}\right).
\label{eq:a_eff}
\end{equation} 
Furthermore, Equation~(\ref{eq:rz_lem}) tells us that we can express $r_z$ as a function of $r_{max}$, allowing us to rewrite Equation~(\ref{eq:a_eff}) in terms of $\chi$;
\begin{equation}
\begin{split}
\alpha_{\text{eff}} &= \arctan\left(\frac{z_{\text{max}}}{2^{-1/2}r_{\text{max}}}\right)\\
&= \arctan\left(\sqrt{2} \chi \right).
\label{eq:a_eff2}
\end{split}
\end{equation}
Using Equation~(\ref{eq:a_eff2}) and the results for the ED model presented above, the $\alpha_{eff}$ for the K-band, KHN-band, and H$\alpha$ line set are estimated ata 1.9$\degr$, 3.0$\degr$, and 14$\degr$, respectively, with a confidence interval ranging from 0$\degr$ to 20$\degr$ from the K-band, 0$\degr$ to 22$\degr$ for the KHN-band, and 0$\degr$ to 32$\degr$ for H$\alpha$. As we can see the ED model predicts larger opening angles than the wedge model; four times greater for H$\alpha$ and up to an order of magnitude greater for the K- and KHN-band sets.

Using the results of the ED model, the radial extent of each emitting region can also be estimated. By combining Equations~(\ref{eq:r_max}) and (\ref{eq:zm_rm}), we can express the radial extent ($r_{max}$) as a function of $\chi$:
\begin{equation}
r_{\text{max}} = \frac{4 \chi^2 v^2}{n}.
\label{eq:r_max2}
\end{equation}

We see that the maximum extent of any region defined by $\chi$ is dependent on the parameter \textit{n} and \textit{v}, meaning $r_{max}$ is also dependent of the stellar and disk parameters as well as its rotational velocity. To compare with previous work in the literature, we decided to adopt a value of 3.5 for \textit{n} (see Section~\ref{sec:theory}) and the stellar parameters of a B1V star (obtained from \cite{cox00}) rotating at $92\%$ of critical angular velocity, the same parameters used by \citet{car11}. Using these parameters, \textit{v} is calculated to be 50.4.

For the K-band and KHN-band set, the maximum extent is estimated to be about 2 to 4 stellar radii. This result closely matches the results of \citet{car11}, who found the K-band emission is contained within 6 stellar radii\footnote{See figure 1 of \citet{car11}.}. For the H$\alpha$ line set, the estimated extent is between 80 and 90 stellar radii, which is slightly larger than the 50 to 60 stellar radii determined by \citet{car11}. One possible cause for these differences could be attributed to a change in the power law $n$. Although the viscous disk model presented above assumes that density structure of the disk is controlled by a constant power law of $n$, some authors have suggested that $n$ might not be constant throughout the disk, but could vary with radius \citep{zor07, car08}. We note, for example, that increasing the value of \textit{n} will result in a smaller extent closer to that determined by \citet{car11}.

\section{DISCUSSION AND CONCLUSION}

For the first time, the geometry of CBe star disks were inferred from the deprojection of axis ratio measurements. A total of 49 ratio measurements from 20 distinct stellar sources collected from the literature were used. These ratios were measured with interferometry in either the K-, H- or N-band, or over the H$\alpha$ emission line. These observations were compared to simulated axis ratios calculated from two disk models; the wedge model, a simple model characterised by an opening half-angle similar to the one proposed by \citet{wat86}, and the equidensity model, whose shape is derived from the viscous disk model. A Monte Carlo technique was employed to generate a large number of simulated ratios, which were compared to the observation by applying Bayesian statistics in order to infer which model best reproduces these observations. 

For the emission regions in the KHN-band, we found that our models can best reproduce the observations with opening half-angles of 0.15$\degr$ to 0.32$\degr$ (wedge model) and 1.9$\degr$ to 3.0$\degr$ (equidensity model) with a confidence interval ranging up to 9.2$\degr$ and 22$\degr$, respectively. Angles of 3.7$\degr$ to 14$\degr$ were found to best reproduce the observation in H$\alpha$ with confidence interval ranging up to 28$\degr$ for the wedge model and 32$\degr$ for the equidensity model. We note that the best fit results are in close agreement with the the opening angles of 2.5$\degr$, 5$\degr$, and 13$\degr$ found by \citet{por96}, \citet{woo97}, and \citet{han96}, respectively, and are below the upper limit of 20$\degr$ determined by \cite{qui97}.

Opening half-angles were also found to be systematically smaller for the KHN-band emission region than the H$\alpha$ region. This also agree with predictions, as the viscous disk model predicts that the scale height of CBe star disk increases with distance from the star \citep{bjo97}. The greater opening angle for H$\alpha$ therefore suggests that its emission region extends to greater distances than the emission region of the KHN-band, which is consistent with the findings of \citet{gie07} and \citet{car11}. 

The extent of the emitting regions for a model star were also estimated, from the results of the ED model, and compared with the results of \citet{car11}. The extent of the KHN-band regions were found to be constrained to a small area close to the star, within 2 to 4 stellar radii. The H$\alpha$ emitting region on the other hand was found to have a much greater area, ranging from 80 to 95 stellar radii. Again, these results agree well with the findings of \citet{car11}, who estimated the emitting regions of the K-band and H$\alpha$ line to be $\sim$5 to 6 and $\sim$50 to 60 stellar radii, respectively.

In this study, a standard distribution of inclination angles (as described in Section~\ref{subsec:dist}) was assumed, without accounting for observational limitations. Due to the limits in resolution power of interferometric measurements, the minor axis of stars seen at high inclination angles (close to equator-on) are less likely to be resolved. This implies that stars with small axis ratios (seen at high inclination) are expected to be under-represented compared to other stars in the sample and that the number of high-inclination stars, and consequently the number of small ratios, in our simulations may be systematically overestimated. 

Our results could also be affected by the visibility models applied to the interferometric observations. As mentioned in Section~\ref{sec:ob}, theoretical models of the visibility of CBe star/disk systems are applied to observations in order to obtain axis ratios. This means that assumptions have already been made on the general shape of the disk and therefore, the results of this study depend on the interferometric models used. 

Finally we note that our models do not take into account the optical thickness of the disk and its effects on the projected axis ratios. Since photons emitted at different locations in the disk have to go through different amounts of material before escaping, light coming from the same equidensity region may not have the same intensity once it reaches us. This could have an effect on our results. To test the significance of this effect, we calculated the optical depth of a disk based on simple isothermal models. For inclinations of 70$\degr$ or lower, we found that the ratios calculated with  differed by no more then 0.02 from the ratios calculated by ignoring optical depth effects, a difference smaller than the uncertainty of the majority of the observations (see Table~\ref{table:ratio}). The effect is more significant for the equator-on case, where we found differences of up to 0.1 between ratios calculated with and without optical depths. Although this difference is larger than seen for lower inclinations, it is still within the order of magnitude of most of the uncertainties in the observed axis ratios.

In conclusion, we found that the results of our deprojections are consistent with the current understanding of CBe star disks. The opening angles were found to be small, supporting the findings of \citet{por96}, \citet{woo97}, and \citet{han96}. We were also able to confirm that H$\alpha$ line emission is formed in a much larger volume of the disk than emission from the KHN-band, as predicted by \citet{car11}. We can therefore conclude that the deprojection method presented in this work can be a very useful tool to obtain information about the size and geometry of CBe star disk based on measured axis ratios. 

\acknowledgments
We thank the anonymous referees for their comments which helped us improve the quality of this paper. We also thank Dr. Rogemar Mamon for his helpful discussions related to statistical methods. C. E. Jones acknowledges the support from NSERC, the Natural Sciences and Engineering Research Council of Canada.

\bibliographystyle{apj}

\end{document}